\documentclass[useAMS,usenatbib]{mn2e}
 \usepackage{graphicx} 
\def\hmpc{~h$^{-1}$ Mpc~}
\def\hkpc{~h$^{-1}$ kpc~}

\title[ Beppo-SAX temperature maps of A2061, A2067, A2124]
{Beppo-SAX temperature maps of galaxy clusters in the Corona
Borealis supercluster: A2061, A2067 and A2124}
\author[F.Marini et al.]{F.~Marini$^{1}$,
S.~Bardelli $^{1}$,\thanks{e-mail: sandro.bardelli@bo.astro.it, 
elena.zucca@bo.astro.it} 
E.~Zucca$^{1}$,\footnotemark[1]
S.~De Grandi$^{2}$,
A.~Cappi$^{1}$,
S.~Ettori $^{3}$,
\newauthor
L.~Moscardini $^{4}$,
G.~Tormen $^{5}$,
A.~Diaferio$^{6}$ \\
$^{1}$ INAF - Osservatorio Astronomico di Bologna via Ranzani 1, 
I--40127 Bologna, Italy\\
$^{2}$ INAF - Osservatorio Astronomico di Brera, 
via Bianchi 46, I--23807 Merate
(LC), Italy \\
$^{3}$ ESO - Karl-Schwarzschild Strasse 2, D--85748 Garching, Germany \\
$^{4}$ Dipartimento di Astronomia, via Ranzani 1, I--40127 Bologna, Italy \\
$^{5}$ Dipartimento di Astronomia, vicolo dell'Osservatorio 2, I--35122 Padova,
Italy \\
$^{6}$ Dipartimento di Fisica Generale ``A. Avogadro'', 
via Giuria 1, I--10125 Torino, Italy 
}

\begin{document}

\date{Accepted 00-00-0000, Received 00-00-0000 in original form}

\pagerange{\pageref{firstpage}--\pageref{lastpage}} \pubyear{2004}

\maketitle

\label{firstpage}

\begin{abstract}
In this paper we present the analysis of Beppo-SAX observations
of the cluster pairs A2061--A2067 and A2122--A2124, located in
the Corona Borealis supercluster, which have been selected as candidate merging
clusters. The aim of this work is to study the physics of
the intracluster medium and to look for the possible presence of merging
signatures.
We derived the global temperatures and abundances and the temperature profiles 
and maps for these clusters.
We do not find any significant evidence of interaction between the clusters 
forming the pairs, but we detect a candidate shock inside A2061. 
On the basis of the X-ray and optical properties of this cluster 
we propose a scenario in which a group is falling inside A2061. 
This interaction is in the phase in which the cores have not
encountered yet and in which the formation of a shock is expected.
\end{abstract}


\section{Introduction}

Clusters of
galaxies are by now recognized to form in a hierarchical way, by the
gravitational merger of smaller clusters and groups. This kind of events are
the most energetic ones in the Universe since the Big Bang, with a release of
$\sim 10^{50}-10^{60}$ erg on a time scale of the order of few Gyrs 
\citep{sarazin00}.
\\
Numerical simulations \citep{tormen97} revealed that mergings happen along 
preferential directions which define matter flow regions, at whose 
intersection rich clusters form. The environment of these intersections is
represented by the core of superclusters: given the high local
density and the small volumes involved, the ``cross section'' for merging
between clusters or group-cluster is strongly enhanced and the dynamical
processes are accelerated. Therefore, the central part of superclusters are
unique and ideal laboratories where to study this astrophysical event.
\\
Applying a percolation algorithm to cluster catalogues, in order to
individuate density enhancements of clusters over the mean density,
\citet{zucca93} obtained catalogues of superclusters at various density 
excesses:
among the richest structures found there are the Shapley Concentration in the
Southern hemisphere and the Corona Borealis supercluster in the Northern one.
\\
Our group already performed an extensive, multiwavelength study of the Shapley
Concentration in the X-ray, optical and radio bands 
\citep{bardelli96,ettori97,bardelli98,venturi00}: in this region three
``cluster complexes'' are found, which present cluster merging events at 
various evolutionary stages.
\\
In order to extend the statistics, we decided to study 
systems of clusters belonging to the central region of the Corona Borealis
supercluster.
\\
In this paper we present the results of Beppo-SAX observations of the cluster
pairs A2061--A2067 and A2122--A2124.
\\
The plan of the paper is the following. In Sect. 2 we introduce the Corona
Borealis supercluster and in Sect. 3 and 4 we describe the data reduction and 
the determination of the temperature profiles and maps, respectively. 
In Sect. 5, 6 and 7 we present the analysis of A2061, A2067, and
A2124 respectively. Finally in Sect. 8 we discuss and summarize the results.
\\
We assume a flat cosmology ($\Omega_{0}$=1) with  $\Omega_{M}$=0.3, 
$\Omega_{\Lambda}$=0.7 and a value for the Hubble constant of 
$H_{0}$=100 km s$^{-1}$ Mpc$^{-1}$.
%
\begin{figure}
\centering
\includegraphics[angle=0,width=\hsize]{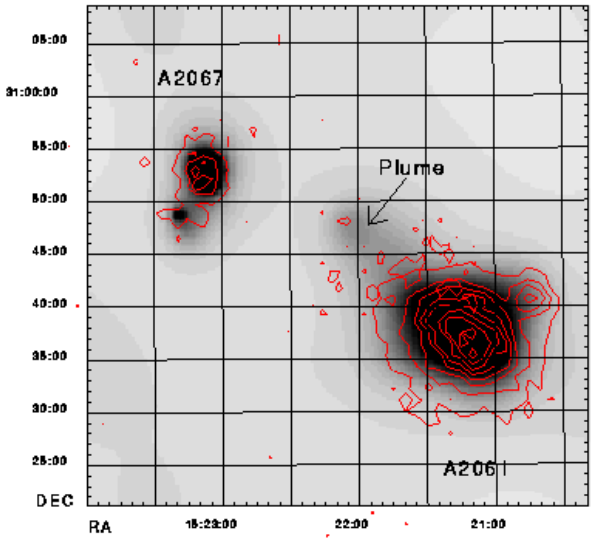}
\caption{X-ray isodensity contours from the Beppo-SAX MECS observations of the 
two clusters A2061 and A2067 superimposed on the image from ROSAT PSPC. The
arrow shows the gas elongation cited in the text.
}
\label{fig:ROSATimage}
\end{figure}
%
\begin{figure}
\centering
\includegraphics[angle=0,width=0.8\hsize]{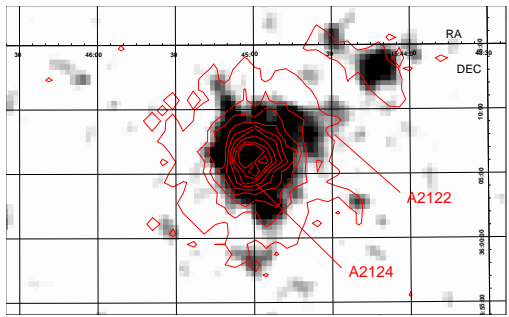}
\caption{X-ray isodensity contours from the Beppo-SAX MECS observation of the 
cluster pair A2122--A2124 superimposed on the Einstein IPC image. The ACO
centres of the two clusters are indicated.
}
\label{fig:Einstein_image}
\end{figure}

\section{The Corona Borealis supercluster}

The Corona Borealis supercluster, at $<z> \sim 0.07$ \citep{postman88},
is the most prominent supercluster in the Northern sky.
\\
At an overdensity in number of clusters of $f \geq 2$ 
\citep{zucca93} this supercluster is made up
of ten clusters: A2019, A2056, A2061, A2065, A2067, A2079, A2089, A2092, A2122 
and A2124 (see also Cappi \& Maurogordato 1992).
\\
The central part of the Corona Borealis supercluster covers a region of
$6^{\circ} \times 6^{\circ}$ centred on $\alpha_{J2000}=15^{h}20^{m}$, 
$\delta_{J2000}=+30^{\circ}$ and is made up of seven clusters: four of them 
(A2056, A2065, A2079 and A2089) are grouped together in the Southern part, 
A2061 and A2067 are close together in the Northern part, and
A2092 is isolated in the North--Eastern part (see also figure 1 in 
Small et al. 1998). 
\\
A2122 and A2124 are located at the North-East periphery of this central region 
($\sim 5^{\circ}$ from the centre of the supercluster).
\\
According to \citet{bahcall92} the entire supercluster extends for at 
least $\sim$ 100 \hmpc on the plane of the sky, while according
to \citet{small97} it is only $\sim$ 40 \hmpc along the line of sight. 
\\
The dominant cluster of Corona Borealis is A2065 ($z$=0.0726, Struble \& Rood 
1999): located in the Southern part of the supercluster, it is the only 
cluster of the supercluster
of richness class 2. It shows a late stage merging \citep{markevitch99}. 
\\
Inside this supercluster two pairs of candidate interacting clusters at high 
density excess have been found: A2061--A2067 ($f$ $\geq 10$) and A2122--A2124 
($f$ $\geq 400$). All these clusters have richness class 1.
\\
In the present work we firstly focus on the clusters A2061 and A2067.
In order to show the relative position of the two clusters, we present
in Figure \ref{fig:ROSATimage} the X-ray isophotes from the two MECS
observations 
in the [2--10] keV energy range superimposed on
a gray scale image of the existing archive ROSAT PSPC pointed observation which
contains both clusters.
Unfortunately, the PSPC observation is offset by $\sim 42$ arcmin, and
therefore the PSF is degradated.
It can be seen from the PSPC observation that A2061 seems to be elongated 
towards A2067. 
This plume of X-ray emission might indicate the presence of interaction between
the two clusters. Note that in the Beppo-SAX data there is no trace of this
elongation because part of it falls under the detector entrance window
supporting structure
(called strongback). From
this image we can estimate that the distance between the centres of the two
clusters on the plane of the sky is $\sim$ 30 arcmin: at the redshift of the 
clusters ($<z> \sim$ 0.076) this distance corresponds to $\sim$ 1.8 \hmpc,
i.e. less than 2 Abell radii
($\sim$ 3 \hmpc). On the other hand the velocity difference 
between A2061 and A2067 is $\Delta v$
$\sim$ 1533 km s$^{-1}$ \citep{oegerle01}: if this $\Delta v$ was due purely 
to Hubble flow, the separation between these two clusters would be 
$\sim$ 15.44 \hmpc, therefore too large for a cluster interaction. 
However this velocity difference results from the composition of the 
unknown relative peculiar velocity and the cosmological distance of
the two clusters.
\\
\begin{table*}
\caption[]{ Beppo-SAX observation log (data refer to MECS).}
\begin{flushleft}
\begin{tabular}{rrrrrrr}
\hline\noalign{\smallskip}
Target & $\alpha$ (2000) & $\delta$ (2000) & date & exp.time & count--rate & 
$n_H$\\
  & $^h$ $^m$ $^s$  & $^o$ $'$ $''$  &   &  ksec  &    cts/s & 
$10^{20}$cm$^{-2}$\\
\noalign{\smallskip}
\hline\noalign{\smallskip}
Abell 2061  & 15 21 16.13 & +30 36 43.93 & 2000 Aug 8--9   & 51.2 & 9.03
$\times 10^{-2}$ &  1.96\\
Abell 2067  & 15 23 08.53 & +30 52 51.33 & 2000 Jul 28--29 & 50.6 & 0.92 
$\times 10^{-2}$  & 1.98\\ 
Abell 2124  & 15 45 00.00 & +36 03 57.60 & 2001 Feb 5--8   & 133.0& 5.19
$\times 10^{-2}$ & 1.65\\ 
\noalign{\smallskip}
\hline
\end{tabular}
\end{flushleft}
\label{tab:logMECS}
\end{table*}
%
Secondly we study the clusters A2122 and A2124.
These clusters are so nearby (their $\Delta v$ is 150 km s$^{-1}$,
while the distance on the plane of the sky is $\sim$ 7 arcmin)
that there are no data about them as pair in
literature: there is a lot of information about A2124, while 
there are very few data about A2122, and sometime they are considered 
as a single cluster. Because
of their small distance on the plane of the sky, they have been observed with
a single Beppo-SAX pointing centred on the cluster A2124.
In Figure \ref{fig:Einstein_image} we present an Einstein IPC image of the
diffuse gas of these
clusters superimposed to the X-ray isophotes from the MECS observation 
in the [2--10] keV energy range: the nominal ACO centres are indicated with
arrows.

\section{Observations and data reduction}

The clusters A2061, A2067 and A2124 were observed by the Beppo-SAX satellite 
\citep{boella97a} in the periods 2000 August 8--9, 2000 July 28--29 and
2001 February 5--8, respectively.
We discuss here the data from two of the instruments on board Beppo-SAX: the
Medium--Energy Concentrator Spectrometer (MECS) and the Low--Energy 
Concentrator Spectrometer (LECS, used here only for A2067). 
The MECS \citep{boella97b} is composed by two units, working in the 
[1--10] keV energy range. At 6 keV, the spectral resolution is $\sim 8\%$ 
and the angular resolution is $\sim 0.7'$ (FWHM). The LECS \citep{parmar97}
consists of an imaging X-ray detector, working in the [0.1--9] keV
energy range, with 20$\%$ spectral resolution and $0.8'$
(FWHM) angular resolution (both computed at 1 keV). Standard
reduction procedures and screening criteria have been adopted to
produce linearized and equalized event files. The MECS (LECS) data
preparation and linearization were performed using the {\sc Saxdas}
({\sc Saxledas}) package under {\sc Ftools} environment.
\\
We also analysed PDS (Phoswich Detector System) data for these clusters: 
we found that no significant detection is present for A2061 and for A2067. 
For A2124 we found a small excess in the [15--40] keV energy range possibly 
associated
with the X-ray source EXSS 1543.1+3615 \citep{oppenheimer97}.
\\
We have taken into account the PSF-induced spectral distortions 
(D'Acri et al. 1998) in the MECS analysis using effective 
area files produced with the {\it effarea} program. Note that, due to
the displacements of the two MECS units, the region masked by the 
support window is different in MECS2 and in MECS3: this fact allows to
recover, at least partly, the information in the region 8--12 arcmin 
from the field center.
A detailed explanation of the MECS analysis is given in \citet{degrandi01}.
All MECS and LECS spectra have been background subtracted using spectra
extracted from blank sky event files in the same region of the
detector as the source \citep{fiore99}.
\\
The observation log is reported in Table \ref{tab:logMECS}. We recall that
A2122 and A2124 belong to a single Beppo-SAX observation. 
The coordinates reported in this table are the pointing coordinates.
The observed
count--rates for A2061, A2067 and A2124 for the 2 MECS units and within 
the central 8 arcmin (corresponding to $\sim$ 0.50 \hmpc) are also reported 
in the table. 
\\
Recently, \citet{perri02} found that a systematic error affects the sky 
coordinates derived from the LECS and MECS instruments. 
The corrections depend on the spacescraft roll angle and are of the order 
of $\sim 30$ arcsec for our observations.
From now on all the X-ray coordinates will be corrected for this systematic 
error. 

\section{Temperature profiles and maps}

In order to find the global temperature, we extracted a circular region
of 8 arcmin ($\sim$ 0.50, $\sim$ 0.48 and $\sim$ 0.40 \hmpc for A2061, A2067
and A2124 respectively) from the centre of the MECS data in the range 
[2--10] keV. This region, although smaller than the extension of the clusters, 
has been chosen because it is not affected by the absorption caused by
the strongback.
We fitted the spectrum with a {\it mekal} \citep{mewe95,kaastra92} model
with an absorbing Galactic hydrogen column 
({\it wabs} model), as implemented in the XSPEC package (Ver. 10.00).     
After having checked that the fitted Galactic absorption is consistent
with the literature measurement \citep{dickey90}, we fixed it to this value.
The values for each cluster are reported in Table \ref{tab:logMECS}.  
\\
In analysing temperature profiles and maps we used only the MECS data,
for which the correction for the PSF-induced spectral distortions is
available.
The cluster emission has been divided into concentric annuli, centred on the 
X-ray emission peak: out to $8$ arcmin the annuli are $2'$ wide, beyond this
radius the annuli are $4'$ wide.
\\
However, we limited all the fitting procedures to the cases in which the source
counts are more than $30\%$ of the total (i.e. source plus background) counts
(see the discussion in De Grandi \& Molendi 2002): indeed when the source
counts are low, the spectral features of the background could dominate the
results.
\\
In order to explore if there is an asymmetry of the temperature distributions
we divided the cluster maps in four sectors. Sector I is North--West of the 
image and the numbers increase clockwise. The value of temperature referred 
to the 0--2 arcmin region is the same for all the sectors because a circular 
region was used to fit data. This analysis has been possible only for
A2061 and A2124, being A2067 too faint to permit meaningful temperature
determinations in sectors.

\begin{figure*}
\centering
\includegraphics[angle=0,width=0.8\hsize]{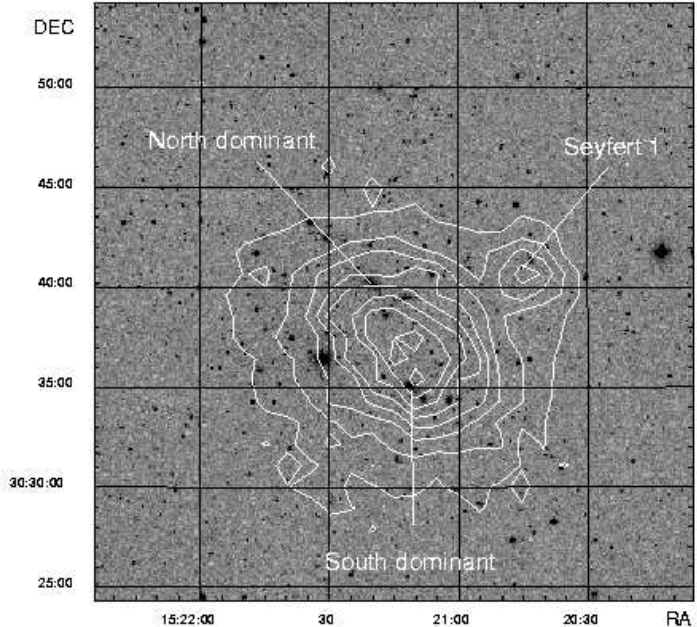}
\caption{X-ray isophotes, from a Beppo-SAX MECS observation pointed on A2061,
superimposed on the optical image from the Digital Sky Survey. The data have 
been smoothed with a Gaussian of 8 pixel FWHM (1 pixel=8 arcsec)
and the linear step within contours is 0.34, with the lowest isophote
corresponding to 0.250 cts pix$^{-1}$. The labelled
objects are described in the text
}
\label{fig:isophotes61}
\end{figure*}

\section{The cluster A2061}

Abell 2061 is a cluster of richness class 1 and Bautz--Morgan class III;
the ACO (Abell et al. 1989) coordinates 
are $\alpha_{2000}=15^{h}21^{m}15^{s}$ and $\delta_{2000}=+30^{h}39'17''$.
\\
On the basis of 118 redshifts \citet{oegerle01} found that the biweight 
estimate of the
observed location was C$_{bi}$ =23699$\pm$ 70 km s$^{-1}$ and that the biweight 
estimate of the scale was S$_{bi}$=780$^{+57}_{-47}$ km s$^{-1}$. 
At the redshift of the cluster 
1 arcmin corresponds to 62.70 h$^{-1}$ kpc.
\\
We observed this cluster using an exposure time of 51163 seconds and 
in Figure \ref{fig:isophotes61} we present the MECS isophotes in the 
[2--10] keV band superimposed on the optical image from the Digital Sky Survey. 
The centroid of diffuse emission has coordinates 
$\alpha_{2000}=15^{h}21^{m}14^{s}$,
$\delta_{2000}=+30^{\circ}37'32''$ and is 1.8 arcmin from the ACO centre, 
corresponding to $\sim$ 0.11 \hmpc.
The hot gas distribution appears to be elongated in the NE--SW direction, 
i.e. toward A2067. 
\\
Two bright galaxies are located approximately
along the major axis of the ellipse described by the isophotes. One of them
(dubbed North dominant) is located 3.1 arcmin ($\sim$ 0.19 \hmpc) 
NE from the X-ray centre; it has a velocity of $v$=23725 $\pm$ 23 km
s$^{-1}$ \citep{oegerle98} and therefore it is at rest with respect to 
the cluster mean velocity. 
The other dominant galaxy (dubbed South dominant), located 2.6 arcmin 
($\sim$ 0.16 \hmpc) SW from the X-ray centre, has a velocity of 
$v$=22720 $\pm$ 20 km s$^{-1}$ \citep{oegerle98}, 
and therefore it has a $\Delta v$
$\sim$ 980 km s$^{-1}$ with respect to the centroid. The R
magnitudes of the two galaxies are R$_{APM}$=12.24 and 12.84, respectively. 
\\
In the NW direction, at 7.9 arcmin ($\sim$ 0.50 \hmpc) from the X-ray centre, 
there is a significant unresolved X-ray peak 
($\alpha_{2000}=15^{h}20^{m}45^{s}$,
$\delta_{2000}=+30^{\circ}40'58.7''$). Near this position two different objects
are located: a galaxy  ($\alpha_{2000}=15^{h}20^{m}45.23^{s}$,
$\delta_{2000}=+30^{\circ}41'0.7''$; R=13.93)
without a measured redshift, at $\sim 20.67$ arcsec from the peak, and a
Seyfert 1 galaxy ($\alpha_{2000}=15^{h}20^{m}43^{s}$,
$\delta_{2000}=+30^{\circ}41'23''$; R=15.27) 
at $\sim 35$ arcsec from the peak, at $z$=0.0772 \citep{small97}, that we
associate to the X-ray peak.

\subsection{Spatial analysis}

Looking at MECS X-ray isophotes of the cluster (see Figure
\ref{fig:isophotes61}), it is clear that it is characterized by an elongated 
diffuse emission. For this reason we fitted the surface brightness distribution 
using an elliptical corrected King law

\begin{eqnarray} 
P(x,y) &=& I_o \left[1+\left({{x'}\over{R_1}} \right)^2 + 
           \left({{y'}\over{R_2}}\right)^2 \right]^{-3\beta + 0.5} +  bck
\end{eqnarray}
\\
where 

\begin{eqnarray*}
 x' &=& (x-x_k) \cos \theta + (y-y_k) \sin \theta    \\
 y' &=& - (x-x_k) \sin \theta + (y-y_k) \cos \theta 
\end{eqnarray*}
\\
in order to take into account the inclination angle $\theta$ of the cluster.
The variables to be estimated are the normalization $I_o$,
the positions ($x_k$, $y_k$) of the centre of the
cluster, the inclination angle of the cluster ($\theta$, computed clockwise
from the North),
its core radii ($R_1$ and $R_2$), the exponent of the King law ($\beta$). The 
background ($bck$) has been estimated from the background map (see Sect. 3) 
and, after having rescaled it by the exposure times, resulted to be 
0.09 cts pix$^{-1}$. 
The fit has been performed on a region of 9 arcmin radius by minimizing the 
$\chi^2$ variable between the model and the data, after having masked a circle
of $24''$ diameter centered on the second X-ray peak described in 
Sect. 5.
The resulting reduced $\chi^2$ is 1.044 with 1254 degrees
of freedom. The results of the fit, done in the [2--10] keV energy range, are
reported in Table \ref{tab:spatial_analysis}.
We found $\beta$=0.634, while the mean value of the core radius 
is 0.265 \hmpc, larger than the average value of 0.08 \hmpc found by
\citet{ettori99} from a sample of 27 clusters. 
\\
This observed large value of the core radius indicates, together with
the other properties described in the previous section, that the cluster 
is not completely relaxed and the ICM is still settling after a major 
disturbance induced from a merger (see the discussion in Sect.8). 
\\ 
\begin{table*}
\caption[]{Results of the bi-dimensional fit for A2061.}
\begin{flushleft}
\begin{tabular}{lllll}
\hline\noalign{\smallskip}
 Normalization  & $R_1$ & $R_2$ & $\beta$ & $\theta$ \\
\noalign{\smallskip}
\hline\noalign{\smallskip}
 $2.199 \pm 0.030$ cts pix$^{-1}$ & $34.97\pm 1.50$ pix & $29.18\pm 1.05$ pix 
        & $0.634\pm 0.023$  & $116^o \pm 0.8^o$ \\
 $(3.13 \pm 0.04) \times 10^{-6}$ erg str$^{-1}$ cm$^{-2}$ s$^{-1}$ &
 $0.290\pm 0.012$ \hmpc & $0.242\pm 0.009$ \hmpc & ~~ & ~~  \\
\hline
\end{tabular}
\end{flushleft}
\label{tab:spatial_analysis}
\end{table*}
%
From the normalization of the King function it is possible to obtain the
central surface brightness, using the appropriate conversion factor from counts
to intrinsic flux (i.e. corrected for absorption). Assuming a bremsstrahlung
emission from a hot gas with temperature of 4.52 keV, 
an abundance of 0.10 and
a hydrogen column density of 1.96 $\times 10^{20}$ atoms cm$^{-2}$
(see Sect. 5.2), the 
central surface brightness in the [2--10] keV range resulted to be
$I_{o}$= 3.13 $\times 10^{-6}$ erg s$^{-1}$ cm$^{-2}$ str$^{-1}$. 
 
\subsection{Global temperature and abundance}
 
\begin{figure}
\centering
\includegraphics[angle=-90,width=0.8\hsize]{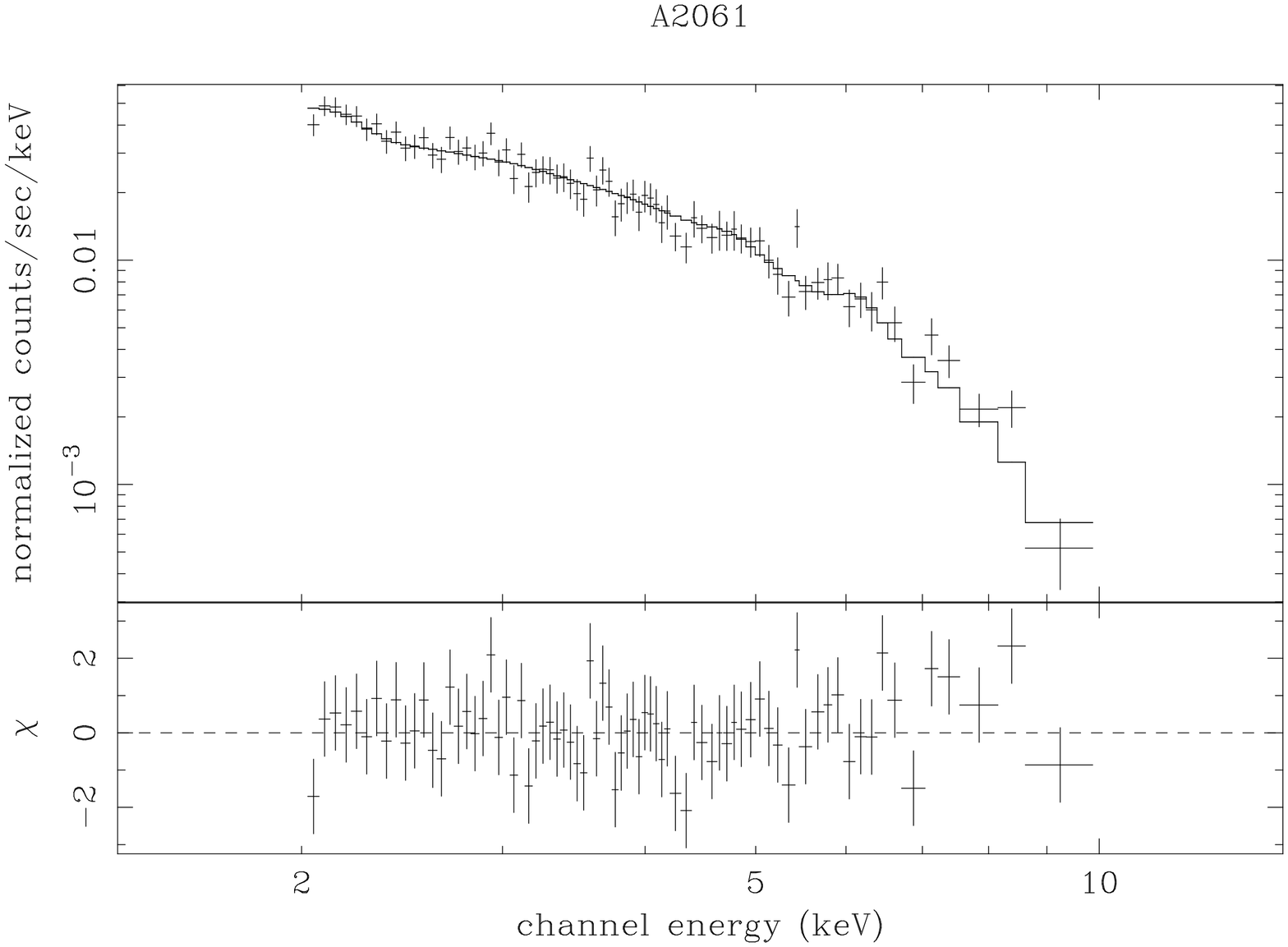}
\includegraphics[angle=-90,width=0.8\hsize]{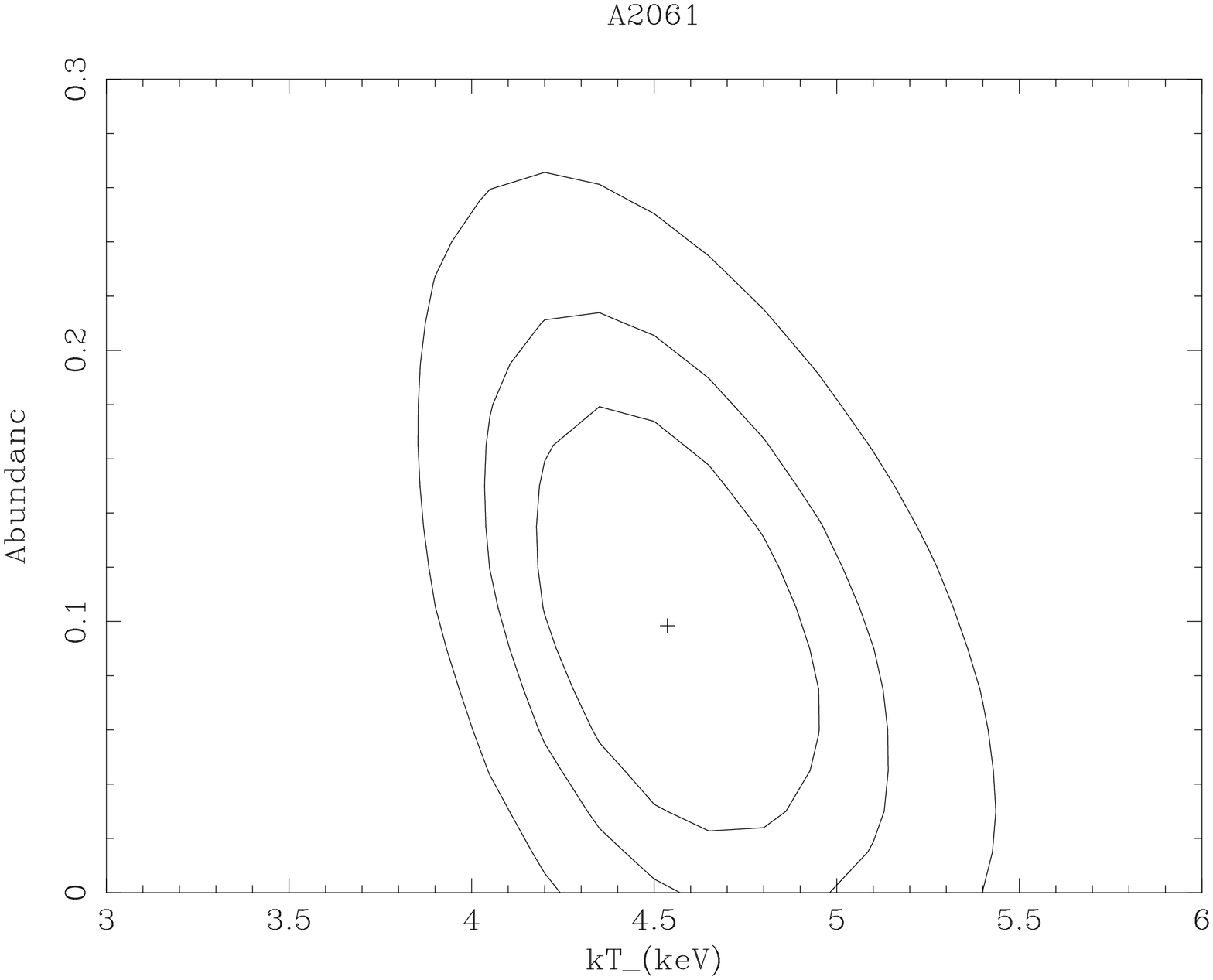}
\caption{Spectrum and confidence ellipse of the estimated temperature
and abundance on MECS data within 8 arcmin from the centre of A2061. The
ellipses correspond to 68\%, 90\% and 99\% confidence levels.}
\label{fig:mecs}
\end{figure}

We found that the global temperature of this cluster, obtained from the
MECS instrument, is $kT=4.52^{+0.48}_{-0.38}$ keV with
an abundance of $0.10^{+0.09}_{-0.08}$,
where the errors are at the $90\%$ confidence level. 
The reduced $\chi^{2}$ is $0.92$ with 109 degrees of freedom. 
In Figure \ref{fig:mecs} we show the MECS spectrum of 
A2061, overplotted to the fit, and the corresponding confidence ellipse 
of the temperature and abundance parameters.
\\
We determined the temperature and the abundance also in a larger 
region (within 16 arcmin, corresponding to $\sim$ 1 \hmpc), finding
$kT=4.13^{+0.44}_{-0.38}$ keV and abundance 
$0.06^{+0.10}_{-0.06}$ with a reduced $\chi^{2}$ of 0.98 and
127 degrees of freedom. These values are well consistent with the previous
estimate. 
\\
%
Following the $\sigma-T$ relation of \citet{lubin93}
[$\sigma=332 \, (kT)^{0.6}$ km s$^{-1}$], the temperature of $kT$=4.52 keV 
implies a velocity dispersion of $821^{+30}_{-26}$ km s$^{-1}$, consistent
with the value estimated by \citet{oegerle01}. 
\\
Another way to estimate the correspondence between the velocity dispersion and
the hot gas temperature is to calculate $\beta_{spec}$  \citep{cavaliere76}:
$$
\beta_{spec} = \frac{\mu m_{p} \sigma^2}{kT}
$$
where $\mu=0.60$ is the mean 
molecular weight, 
$m_{p}$ is the proton mass, $\sigma$ is the velocity
dispersion of galaxies and $kT$ is the temperature of the cluster gas. 
Using the value obtained by Oegerle et al. for the velocity dispersion, 
and the temperature we found, we obtained 
$\beta_{spec}$=0.84, lower than the typical value of 1.1 \citep{sarazin88}. 
\\
With the temperature estimated in the region 0--16 arcmin, the 
total luminosity within this region
is $L_{[2-10]keV}=(1.21\pm 0.02) \times 10^{44}$ h$^{-2}$ erg s$^{-1}$, 
corresponding to a bolometric luminosity of  
$L_X$=2.45 $\times 10^{44}$ h$^{-2}$ erg s$^{-1}$;
the flux we obtained in the same region is $F_{[2-10]keV}=(1.72 \pm 0.03)
\times 10^{-11}$ erg cm$^{-2}$ s$^{-1}$. 
We found these two values after having
subtracted the contribution of the second X-ray peak described in Sect. 5.
\\
\citet{ebeling96} reported the flux for this cluster from the ROSAT All
Sky Survey, assuming a temperature of 5.5 keV: they found $F_{[2-10]keV}$=1.46 
$\times 10^{-11}$ erg cm$^{-2}$ s$^{-1}$, i.e. 15\% lower than our value.
From the X-ray luminosity function of clusters in the [2--10] keV band, 
\citet{edge90} found $L^{*}$=2 $\times 10^{44}$ h$^{-2}$ erg 
s$^{-1}$: the luminosity we estimated is 39\% lower than $L^{*}$.

\subsection{Temperature profiles and maps}

\begin{figure}
\centering
\includegraphics[angle=0,width=\hsize]{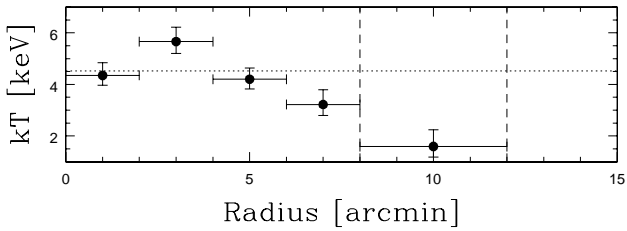}
\caption{Temperature radial profile of A2061. The vertical bars correspond to
68\% errors, while the horizontal bars represent the bins used to extract the
counts. The dotted line corresponds to the global temperature, derived within 
8 arcmin; vertical dashed lines indicate the region where the strongback 
correction is applied. 
}
\label{fig:Tprofile61}
\end{figure}

The next step is the description of the temperature distribution within
the cluster by dividing the frame in rings and sectors.
Given the low count statistics, the abundance resulted not constrained by 
the data, so we fixed it to the 
value of 0.10, according to the value found with MECS data within 8 arcmin. 
The spectral analysis was performed after having masked a circle
of $3'$ radius centered on the second X-ray peak described in 
Sect. 5. This peak affects annuli 6$'$--8$'$ and 8$'$--12$'$. 
The results are reported in Table \ref{tab:spectralannuli}.
Figure \ref{fig:Tprofile61} shows the temperature profile of A2061
in annuli around the cluster centre. The vertical bars correspond to the 
$68\%$ errors and the horizontal bars represent the bins used to extract the 
counts. The dotted line corresponds to the value obtained from the global 
temperature fit.
From this figure we note that the profile is slightly decreasing, a part the
second point, which is higher than the global value: this is due to a
temperature increase in a specific area of the cluster (see below).
\\
\begin{table}
\caption[]{Spectral results from regions of various radii (arcmin) from the
centre of
A2061. Errors are at 90 \% and 68 \% confidence level. Abundance is fixed to the
value 0.10.}
\begin{flushleft}
\begin{tabular}{rrr}
\hline\noalign{\smallskip}
Radius ($arcmin$)& kT ($keV$) &  Reduced $\chi^{2}$ (d.o.f.)\\
\noalign{\smallskip}
\hline\noalign{\smallskip}
0 - 2  & $4.35^{+0.87 \; +0.49}_{-0.61 \; -0.38}$
       & $0.64 \; (48)$ \\  
2 - 4  & $5.66^{+0.98 \; +0.56}_{-0.74 \; -0.46}$
       & $0.79 \; (58)$ \\
4 - 6  & $4.20^{+0.84 \; +0.44}_{-0.64 \; -0.38}$
       & $1.11 \; (50)$ \\
6 - 8  & $3.22^{+1.03 \; +0.57}_{-0.67 \; -0.43}$
       & $1.02 \; (27)$ \\
8 - 12 & $1.60^{+1.31 \; +0.65}_{-0.62 \; -0.41}$
       & $0.90 \; (38)$ \\
\noalign{\smallskip}
\hline
\end{tabular}
\end{flushleft}
\label{tab:spectralannuli}
\end{table}
%
%
\begin{figure}
\centering
\includegraphics[angle=0,width=\hsize]{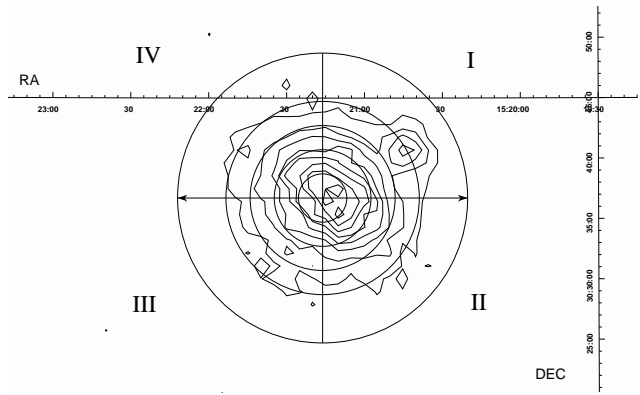}
\caption{Beppo-SAX MECS image of A2061. The concentric circles correspond to the
bins of the radial profiles, while the quadrants correspond to the sectors used
for the temperature map analysis.
}
\label{fig:sector}
\end{figure}
%
\begin{table}
\caption[]{Spectral results from regions of various radii (arcmin) from the
centre of
A2061. Errors are at 90 \% and 68 \% confidence level. Abundance is fixed to the
value 0.10. Sectors indicated in italic are those where the source counts are
lower than 30\% of the total.}
\begin{flushleft}
\begin{tabular}{rrrr}
\hline\noalign{\smallskip}
R ($arcmin$)& Sector & kT ($keV$) & Reduced $\chi^{2}$ (d.o.f.)\\
\noalign{\smallskip}
\hline\noalign{\smallskip}
2 - 4  & I
       & $4.27^{+1.45 \; +0.80}_{-0.95 \; -0.62}$
       & $0.65 \; (46)$ \\ 
~~~~~~ & II
       & $4.65^{+1.45 \; +0.81}_{-0.95 \; -0.61}$
       & $0.45 \; (49)$ \\
~~~~~~ & III
       & $4.78^{+2.19 \; +1.14}_{-1.20 \; -0.78}$
       & $0.41 \; (37)$ \\ 
~~~~~~ & IV
       & $10.67^{+8.35 \; +3.90}_{-3.60 \; -2.47}$
       & $0.66 \; (44)$ \\  
\noalign{\smallskip}
\hline\noalign{\smallskip}                  
4 - 6  & I
       & $2.91^{+1.29 \; +0.72}_{-0.71 \; -0.45}$
       & $0.57 \; (46)$ \\ 
~~~~~~ & II
       & $5.23^{+2.65 \; +1.39}_{-1.46 \; -0.98}$
       & $0.64 \; (34)$ \\
~~~~~~ & III
       & $5.05^{+5.33 \; +2.40}_{-2.00 \; -1.40}$
       & $0.69 \; (49)$ \\ 
~~~~~~ & IV
       & $3.96^{+1.40 \; +0.73}_{-0.90 \; -0.60}$
       & $0.58 \; (24)$ \\
\noalign{\smallskip}
\hline\noalign{\smallskip}                    
6 - 8  & I
       & $4.53^{+20.60 \; +9.25}_{-2.64 \; -1.83}$
       & $0.36 \; (57)$ \\ 
~~~~~~ & II
       & $3.80^{+3.76 \; +1.75}_{-1.51 \; -1.07}$
       & $0.40 \; (39)$ \\
~~~~~~ & III
       & $2.50^{+1.69 \; +0.86}_{-0.84 \; -0.55}$
       & $0.79 \; (06)$ \\ 
~~~~~~ & IV
       & $2.39^{+2.29 \; +1.08}_{-1.10 \; -0.76}$
       & $0.31 \; (36)$ \\ 
\noalign{\smallskip}
\hline\noalign{\smallskip}                   
8 - 12 & {\it I}
       & ${\it 2.55^{+20.50 \; +3.45}_{-1.65 \; -1.05}}$
       & ${\it 0.41 \; (11)}$ \\
~~~~~~ & {\it II}
       & ${\it 2.13^{+1.60 \; +0.80}_{-0.81 \; -0.55}}$
       & ${\it 1.10 \; (13)}$ \\
~~~~~~ & {\it III}
       & ${\it 1.08^{+1.27 \; +0.64}_{-0.49 \; -0.33}}$
       & ${\it 1.53 \; (12)}$ \\
~~~~~~ & IV
       & $2.77^{+1.19 \; +0.63}_{-0.75 \; -0.49}$
       & $1.10 \; (18)$ \\           
\noalign{\smallskip}
\hline
\end{tabular}
\end{flushleft}
\label{tab:sector61}
\end{table}
%
\begin{figure}
\centering
\includegraphics[angle=0,width=\hsize]{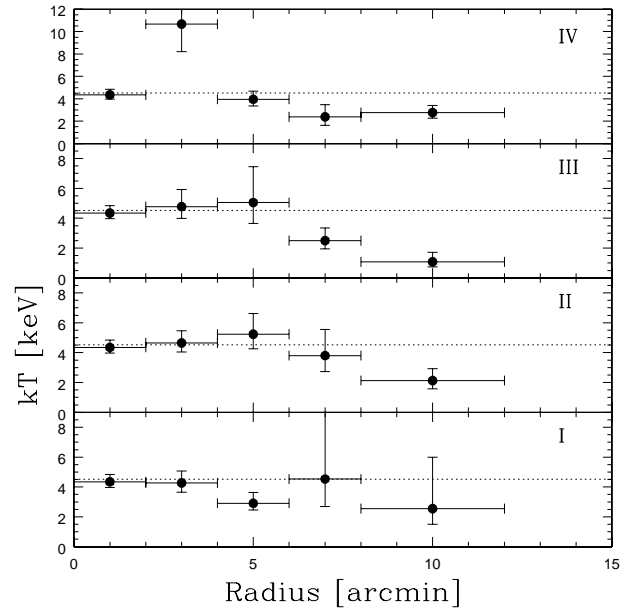}
\caption{Temperature map of A2061. The vertical bars correspond to
68\% errors, while the horizontal bars represent the bins used to extract the
counts. Dotted lines correspond to the global temperature fit.
For the first bin a circular region was used to fit data, consequently the
value is the same for all quadrants.}
\label{fig:sectorTprofile61}
\end{figure}
%
\begin{figure}
\centering
\includegraphics[angle=-90,width=0.8\hsize]{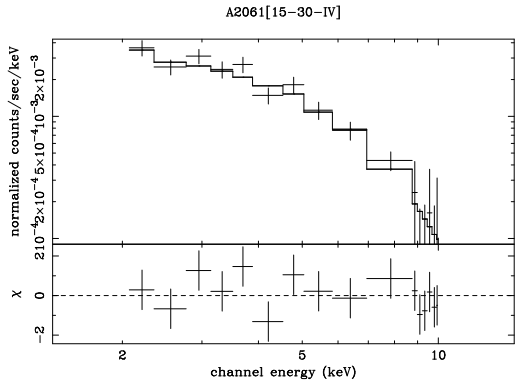}
\caption{Spectrum, fit and residuals of the MECS data in the 2$'$--4$'$ arcmin
annulus, sector IV, for A2061.
}
\label{fig:spectrum_15_30_IV}
\end{figure}
%
In order to explore if there is an asymmetry of the temperature distribution
due to the interaction of the intracluster medium with possible 
material of the A2067 cluster (see Figure \ref{fig:ROSATimage}), we divided 
the cluster map in four sectors as shown in Figure \ref{fig:sector}.  
Sector IV points towards the A2067 cluster.
Also in this case, we masked the second X-ray peak, which is located in 
sector I.
The results are reported in Table \ref{tab:sector61}. 
Note that in the 8$'$--12$'$
annuli, sectors I, II and III, the source counts are 20.0 \%, 17.8\% and 18.7\%,
respectively, of the total counts, that is the background
contribution becomes dominant.
In Figure \ref{fig:sectorTprofile61} the temperature profiles in the four 
sectors are presented. 
Looking at the image it is possible to note that 
the temperature distribution is quite isothermal up
to 8 arcmin, with the exception of sector IV, for which there is a remarkable
increase of the temperature around the 3 arcmin bin which is not consistent 
(at 2.5 sigma) with that of the global fit. This result leads us to suppose 
that the
discrepancy with respect to the global value found in the temperature radial 
profile in the 2$'$--4$'$ arcmin annulus (see Figure \ref{fig:Tprofile61}) is 
due to the presence of this excess. 
\\
The spectral fit of the data in the 2--4 region of sector IV is shown in
Figure \ref{fig:spectrum_15_30_IV}: the temperature is 
$kT=10.67^{+3.90}_{-2.47}$
keV (the errors are at the 68\% confidence level) and the reduced $\chi^2$ is
0.96 with 44 degrees of freedom. 
\\
In order to check the robustness of the temperature determined in this region,
we repeated the fit,
forcing the abundance to different values in the range 0.00--1.00: the derived 
temperatures are in the range
9.06--11.45 keV, in any case not consistent with the value from the global fit. 
Moreover we checked whether the data could be described by a bremsstrahlung 
plus a power law: the fit, although formally acceptable, does not constraint the
photon index of the power law.
\\
This result might suggest the presence of a candidate  internal shock due to
the merger of an infalling group into the main body of the cluster, as we
discuss in Sect. 8.

\begin{figure*}
\centering
\includegraphics[angle=0,width=0.8\hsize]{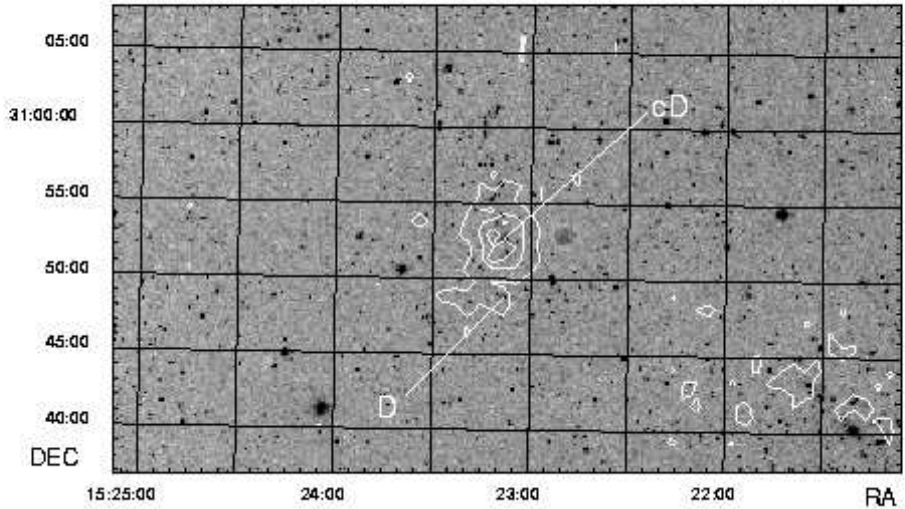}
\caption{X-ray isophotes, from a Beppo-SAX MECS observation pointed on A2067,
superimposed on the optical image from the Digital Sky Survey. The data have 
been smoothed with a Gaussian of 6 pixel FWHM (1 pixel=8 arcsec)
and the linear step within contours is 0.22, with the lowest isophote
corresponding to 0.250 cts pix$^{-1}$. The labelled
objects are described in the text.
}
\label{fig:isophotes67}
\end{figure*}

\section{The cluster A2067}

Abell 2067 is a cluster of richness class 1 and Bautz-Morgan class III; its
centre has coordinates $\alpha_{2000}=15^{h}23^{m}14^{s}$,
$\delta_{2000}=+30^{\circ}54'23''$ \citep{aco}.
On the basis of 44 redshifts,  \citet{oegerle01} estimated 
C$_{bi}$=22166$\pm$ 79 km s$^{-1}$ and 
S$_{bi}$=536$^{+69}_{-90}$ km s$^{-1}$.
At the redshift of the cluster  
1 arcmin corresponds to 59.98 h$^{-1}$ kpc.
\\
In Figure \ref{fig:isophotes67} we present the MECS X-ray isophotes superimposed
on the optical image from the Digital Sky Survey: the isodensity contours of the
cluster are taken in the energy range [2--10] keV.
The centroid of diffuse emission has coordinates
$\alpha_{2000}=15^{h}23^{m}08^{s}$,
$\delta_{2000}=+30^{\circ}53'05''$ and is 1.9 arcmin from the ACO centre, 
corresponding to $\sim$ 0.11 \hmpc. 
This cluster does not appear significantly elongated but shows a diffuse
excess in the Southern part, in correspondence of a galaxy overdensity.
\citet{kopylov98} reported that A2067 is
characterized by the presence of a central dominant galaxy 
($\alpha_{2000}=15^{h}24^{m}00^{s}$, $\delta_{2000}=+31^{\circ}12'40''$),
but we found that this object is 22.6 arcmin (1.36 \hmpc) far from the 
X-ray centre.
Studying the morphological classification of galaxies which belong to A2067 we
found out two dominant galaxies: one of them, a central dominant galaxy, is
located 0.5 arcmin ($\sim$ 0.03 \hmpc) from the X-ray centre,
has a velocity of $v$=22061 $\pm$ 54 km s$^{-1}$ \citep{postman88}
and its R$_{APM}$ magnitude is 12.44.
The other dominant galaxy, with a velocity of 
$v$=21576 $\pm$ 50 km s$^{-1}$ \citep{postman88} is located 4.3 arcmin 
($\sim$ 0.26 \hmpc) from the X-ray centre ($\alpha_{2000}=15^{h}23^{m}08^{s}$,
$\delta_{2000}=+30^{\circ}48'46''$) and its R$_{APM}$ magnitude is 15.97. 
These dominant galaxies are indicated as cD and D in Figure
\ref{fig:isophotes67}.

\subsection{Global temperature and abundance}

\begin{figure}
\centering
\includegraphics[angle=-90,width=0.8\hsize]{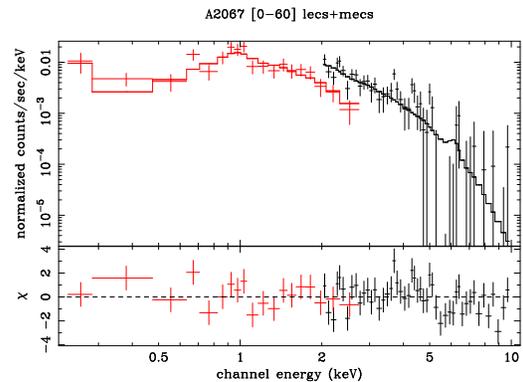}
\caption{Spectrum of LECS$+$MECS data within 8 arcmin from the centre of A2067.}
\label{fig:lecsmecs67}
\end{figure}

Fitting the data 
extracted from a circular region of 8 arcmin from the centre of the MECS data in
the range [2--10] keV we found  
$kT=1.73^{+0.38}_{-0.37}$ keV (errors are at the 90\% confidence level) 
with a reduced $\chi^2$ of 1.52 with 46 degrees of freedom, but the abundance is
not constrained. Therefore in order to use all the information present in the 
Beppo-SAX observation, we considered also the data of the LECS instrument in 
the [0.2--3] keV band extracted from the same region: we obtained 
$kT=1.44^{+0.38}_{-0.30}$ keV and an abundance of $0.18^{+0.42}_{-0.16}$, 
with a reduced $\chi^2$ of 1.08 with 16 degrees of freedom.
The temperature is consistent with the MECS value.
\\
We have repeated the fit with the MECS data in the region 0--8 arcmin fixing the
abundance to the value found from LECS data: we obtained
$kT=1.67^{+0.37}_{-0.31}$ keV  
with a reduced $\chi^2$ of 1.50 with 47 degrees of freedom.
\\
Finally we used the combined LECS$+$MECS data.
In this case, we ran the fits by estimating also the relative normalization 
between the two instruments: we obtained a value similar to that found by 
\citet{bardelli02} and by \citet{ettori00}. Therefore we fixed it to the
value 0.50.
From the LECS$+$MECS data we found $kT=1.54^{+0.26}_{-0.22}$ keV and an 
abundance of $0.19^{+0.30}_{-0.16}$, 
with a reduced $\chi^2$ of 1.39 with 64 degrees of freedom. The temperature 
is consistent at 0.6 sigma with that obtained from MECS data and at
0.4 sigma with that obtained from LECS data; the abundance is consistent at 0.1
sigma with the value found out from LECS data.
In Figure \ref{fig:lecsmecs67} we show the combined LECS$+$MECS spectrum of 
A2067, overplotted to the fit. 
\\
The temperature of $kT$=1.54 keV implies a velocity dispersion of 
$430^{+26}_{-24}$ km s$^{-1}$ (errors are at 68\% confidence level) which is
consistent with the value estimated by \citet{oegerle01}. 
\\
From MECS data we found that the total luminosity within 8 arcmin is 
$L_{[2-10]keV}=(0.58\pm 0.04) \times 10^{43}$ h$^{-2}$ erg s$^{-1}$, 
corresponding to a bolometric luminosity of 
$L_X$=2.11 $\times 10^{43}$ h$^{-2}$ erg s$^{-1}$.
According to \citet{mckee80}, the X-ray luminosity in the range [2--10] keV of 
A2067 is $L_{[2-10]keV} <$ 1.31 $\times 10^{43}$ h$^{-2}$ erg s$^{-1}$.
Our value is $\sim$ 97\% lower than the $L^{*}$ value given by \citet{edge90}.

\begin{table}
\caption[]{Spectral results from regions of various radii (arcmin) from the
centre of
A2067. Errors are at 90 \% and 68 \% confidence level. Abundance is fixed to the
value 0.18. Regions indicated in italic are those where the source counts are
lower than 30\% of the total.}
\begin{flushleft}
\begin{tabular}{rrr}
\hline\noalign{\smallskip}
Radius ($arcmin$)& kT ($keV$) & Reduced $\chi^{2}$ (d.o.f.)\\
\noalign{\smallskip}
\hline\noalign{\smallskip}
0 - 2  & $1.96^{+0.41 \; +0.24}_{-0.32 \; -0.20}$
       & $1.91 \; (13)$ \\ 
2 - 4  & $1.28^{+0.49 \; +0.28}_{-0.34 \; -0.22}$
       & $1.22 \; (11)$ \\
{\it 4 - 6}  & ${\it 1.76^{+1.41 \; +0.68}_{-0.74 \; -0.50}}$
       & ${\it 1.16 \; (15)}$ \\
\noalign{\smallskip}
\hline
\end{tabular}
\end{flushleft}
\label{tab:spatial67}
\end{table}

\subsection{Temperature profiles}

\begin{figure}
\centering
\includegraphics[angle=0,width=\hsize]{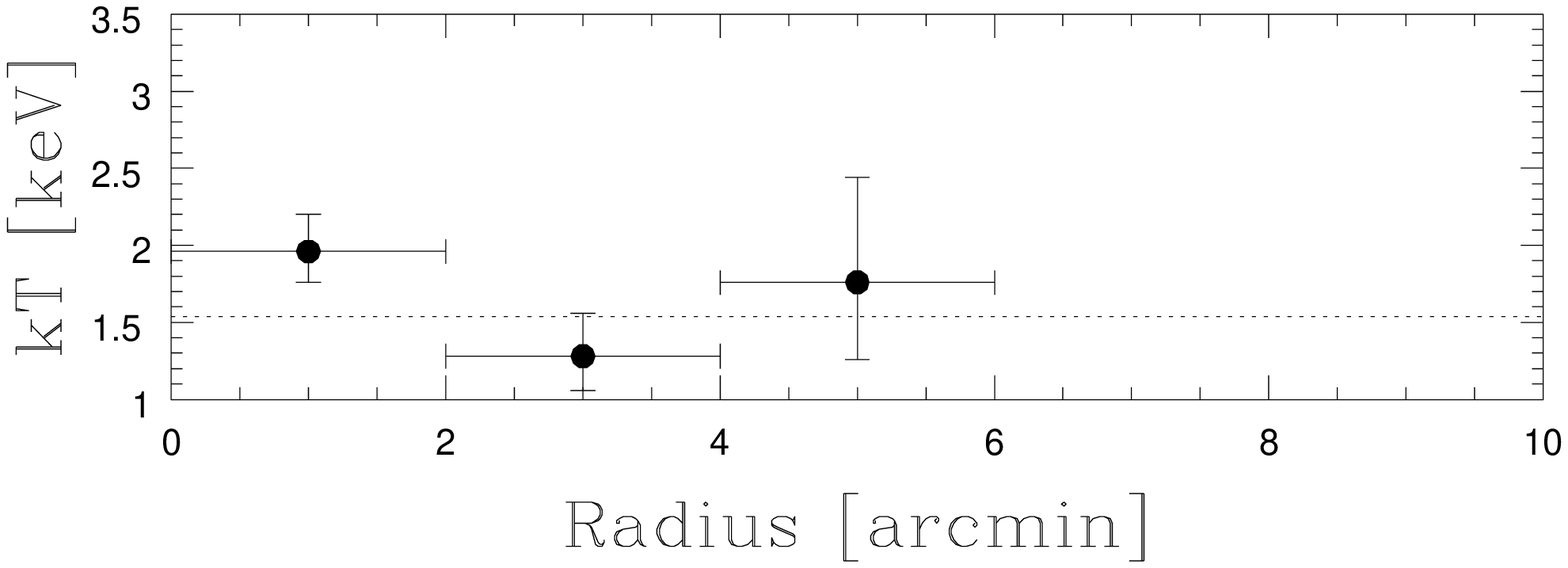}
\caption{Temperature radial profile of A2067. The vertical bars correspond to
68\% errors, while the horizontal bars represent the bins used to extract the
counts. The dotted line corresponds to the global temperature fit, derived
within 8 arcmin.
}
\label{fig:Tprofile67}
\end{figure}

In analysing temperature profiles we fixed the abundance 
to the value of 0.18, according to the value found with LECS data. 
The results are reported in Table \ref{tab:spatial67}. Note that in the
4$'$--6$'$
annulus the source counts are 27.1 \% of the total counts, i.e. lower than 30\%
of the total and therefore on the border of our acceptability criterium.
In Figure \ref{fig:Tprofile67} we report the temperature profile of A2067
in annuli around the cluster centre. 
Looking at the figure it is possible to note that the temperature is roughly
constant, i.e. it is isothermal up to 6 arcmin ($\sim$ 0.36 \hmpc).
\\
Given the low statistics of counts we were not able to divide the cluster in
sectors as done for A2061 and it was not possible to do a bidimensional
analysis. 

\section{The cluster A2124}

\begin{figure*}
\centering
\includegraphics[angle=0,width=0.8\hsize]{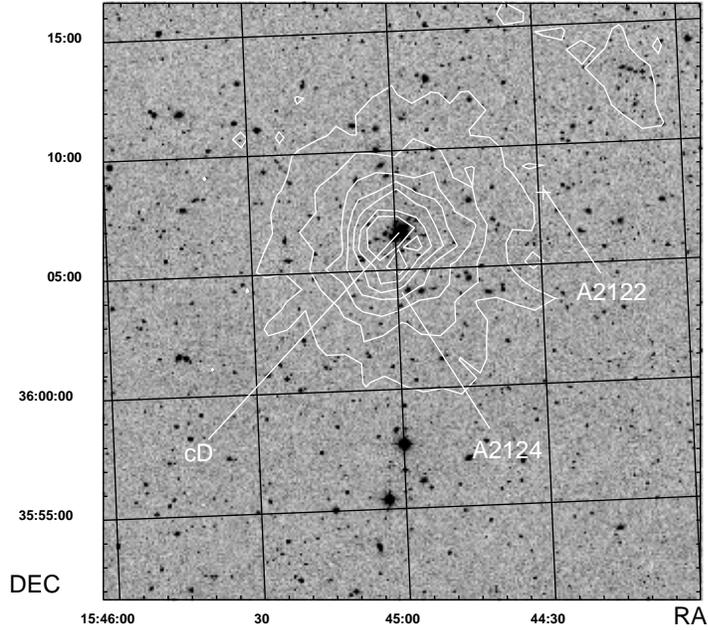}
\caption{X-ray isophotes, from a Beppo-SAX MECS observation, of the cluster
A2124,
superimposed on the optical image from the Digital Sky Survey. The data have 
been smoothed with a Gaussian of 6 pixel FWHM (1 pixel=8 arcsec)
and the linear step within contours is 0.50, with the lowest isophote
corresponding to 0.571 cts pix$^{-1}$. The labelled
objects are described in the text.
}
\label{fig:isophotes2124}
\end{figure*}

Abell 2124 is a richness class 1 cluster and Bautz-Morgan class I; its
centre has coordinates $\alpha_{2000}=15^{h}44^{m}59^{s}$,
$\delta_{2000}=+36^{\circ}04'$ \citep{aco}.
On the basis of 61 redshifts \citet{oegerle01} estimated 
C$_{bi}$=19684$\pm$ 110 km s$^{-1}$ and 
S$_{bi}$=862$^{+91}_{-69}$ km s$^{-1}$.
At the redshift of the cluster 1 arcmin corresponds to 52.92 h$^{-1}$ kpc.
\\
We observed this cluster with an exposure time of 133017 seconds.
In Figure \ref{fig:isophotes2124} we present the MECS X-ray isophotes 
superimposed on the optical image from the Digital Sky Survey.
The X-ray centre has coordinates $\alpha_{2000}=15^{h}44^{m}56^{s}$,
$\delta_{2000}=+36^{\circ}06'04''$.
A2124 is dominated by a single cD galaxy ($\alpha_{2000}=15^{h}44^{m}59^{s}$,
$\delta_{2000}=+36^{\circ}06'34''$), with a velocity of $v$=19810 km s$^{-1}$  
and R magnitude of 12.49 \citep{kopylov98}.
\citet{blakeslee99} identified an arclike object at $z$=0.573
located 27 arcsec along the major axis from the centre of this cD galaxy.
\\
As discussed in Sect. 2., in this Beppo-SAX observation 
another Abell cluster (A2122) is included with a position
NW from A2124 (in Figure \ref{fig:isophotes2124} 
we indicate the ACO centre of both clusters).
Abell 2122 is reported as a cluster of richness 1 and Bautz-Morgan 
class II-III; its
centre has coordinates $\alpha_{2000}=15^{h}44^{m}29^{s}$,
$\delta_{2000}=+36^{\circ}08'$ \citep{aco}
and its velocity is $v$=19830 km s$^{-1}$ \citep{struble99}. 
There is apparently a lot of confusion in the literature about A2122 and A2124.
These two clusters are at a projected separation of 7.3 arcmin. 
\citet{abell58}
classified both of them as richness 1 clusters, but assigned distance class 3
to A2122 and distance class 5 to A2124. Looking at the DSS images centred on
the two clusters, A2124 appears indeed as a poorer concentration of fainter
galaxies 3 arcmin South to the cD (but at the same right ascension). However,
the centre of A2122 falls West of the cD, in a region which apparently does not
correspond to any significant galaxy concentration. Quite confusingly, in the 
NED database only A2122 is associated to a Zwicky cluster, while the SIMBAD 
database gives wrong
coordinates for A2122, $\alpha_{2000}= 15^h 44^m 48^s$ and $\delta_{2000} 
= 36^o 06' 00''$ (instead of $\alpha_{2000} = 15^h 44^m 29^s$ and 
$\delta_{2000}=36^o 07' 38''$, as reported in NED).
The available redshift measurements show that the two clusters are at the same 
distance, at $z \sim 0.066$: this means that the separation of 7.3 arcmin 
corresponds to only
$\sim 0.39$ h$^{-1}$ Mpc; on this basis and taking into
account the X--ray map, we conclude that A2122 and A2124 should be 
identified with the same cluster.

\begin{table*}
\caption[]{Results of the bi-dimensional fit for A2124.}
\begin{flushleft}
\begin{tabular}{lllll}
\hline\noalign{\smallskip}
 Normalization  & $R_1$ & $R_2$ & $\beta$ & $\theta$ \\
\noalign{\smallskip}
\hline\noalign{\smallskip}
 $5.189\pm 0.050$ cts pix$^{-1}$ & $18.38 \pm 0.50$ pix & $16.13 \pm 0.65$ pix
               & $0.586\pm 0.010$  & $95^o \pm 2^o$\\
 $(2.91\pm 0.03) \times 10^{-6}$ erg str$^{-1}$ cm$^{-2}$ s$^{-1}$ &
 $0.130\pm 0.004$ \hmpc & $0.114 \pm 0.005$ \hmpc & ~~ & ~~  \\
\hline
\end{tabular}
\end{flushleft}
\label{tab:spatial_analysis_A2124}
\end{table*}
%

\subsection{Spatial analysis}

As done in Sect. 5.1., we fitted the surface brightness distribution 
using an elliptical King law; 
the background ($bck$) has been estimated from the background map, and resulted 
to be 0.205 cts pix$^{-1}$. 
The resulting reduced $\chi^2$ is 0.956 with 1254 degrees of freedom. 
The results of the fit, done in the [2--10] keV energy range and within a
region of 9 arcmin radius, are
reported in Table \ref{tab:spatial_analysis_A2124}.
We found $\beta$=0.586, while the average value of the core radius 
is 0.122 \hmpc.
\\
With these values and the temperature profile (see Sect.7.3) we derive
a total mass within 0.45 \hmpc of $1.11 \times 10^{14}$ h$^{-1}$ M$_\odot$.

\subsection{Global temperature and abundance}

For the MECS global temperature we found $kT$=4.41$^{+0.38}_{-0.33}$ keV and 
an abundance of 0.29$^{+0.10}_{-0.09}$, where the errors are
at the 90\% confidence level. The reduced $\chi^{2}$ is 0.98 with 132 degrees
of freedom.
\\    
The temperature found implies a velocity dispersion
of 809$^{+24}_{-23}$ km s$^{-1}$ (errors are at 68\% confidence level), which
is consistent with the value estimated by \citet{oegerle01}.
\\
Using the value obtained by Oegerle for the velocity dispersion, 
and the temperature we found from MECS data in the 0-8
arcmin region, we obtained $\beta_{spec}$=1.05, 
similar to the typical value of 1.1 \citep{sarazin88}. 
\\
From MECS data we found that the total luminosity within 8 arcmin is 
$L_{[2-10]keV}=(2.82\pm 0.04) \times 10^{43}$ h$^{-2}$ erg s$^{-1}$, 
corresponding to a bolometric luminosity of  
$L_X$=0.56 $\times 10^{44}$ h$^{-2}$ erg s$^{-1}$;
the flux we obtained in the same region is 
$F_{[2-10]keV}=(0.58\pm 0.01) \times 10^{-11}$ erg cm$^{-2}$ s$^{-1}$ 
to be compared with 
$F_{[2-10]keV}$=0.48 $\times 10^{-11}$ erg cm$^{-2}$ s$^{-1}$, estimated 
by \citet{ebeling96}.
The luminosity we estimated is 86\% lower than the $L^{*}$ value 
given by \citet{edge90}.

\subsection{Temperature profiles and maps}

\begin{figure}
\centering
\includegraphics[angle=0,width=\hsize]{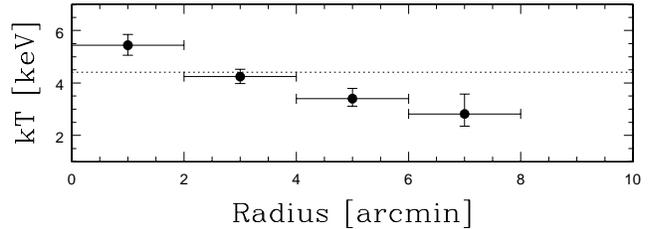}
\caption{Temperature radial profile of A2124. The vertical bars correspond to
68\% errors, while the horizontal bars represent the bins used to extract the
counts. The dotted line corresponds to the global temperature fit, derived 
within 8 arcmin.
}
\label{fig:Tprofile24}
\end{figure}

\begin{figure}
\centering
\includegraphics[angle=0,width=\hsize]{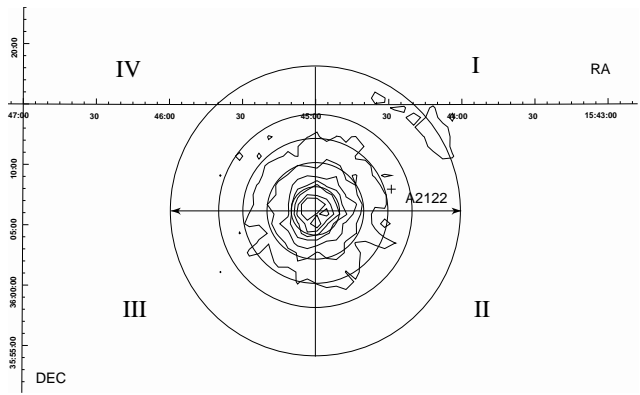}
\caption{Beppo-SAX MECS image of A2124. The concentric circles correspond to the
bins of the radial profiles, while the quadrants correspond to the sectors used
for the temperature map analysis. The small cross indicates the ACO centre of
A2122.
}
\label{fig:sector_24}
\end{figure}

\begin{figure}
\centering
\includegraphics[angle=0,width=\hsize]{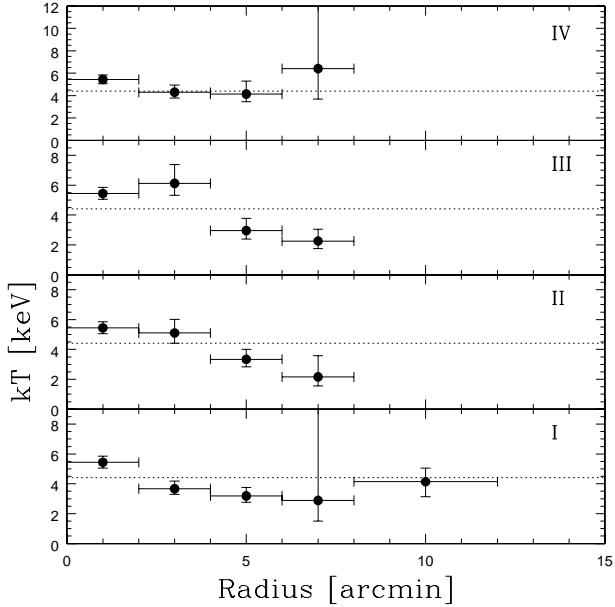}
\caption{Temperature map of A2124. The vertical bars correspond to
68\% errors, while the horizontal bars represent the bins used to extract the
counts. Dotted lines correspond to the global temperature fit.
For the first bin a circular region was used to fit data.}
\label{fig:sectorTprofile24}
\end{figure}

In analysing temperature profiles we used the MECS data,
and divided the cluster emission in the same way as done for A2061 and A2067 
(see Sect. 5.3 and 6.2).
Abundance is fixed to the
value of 0.29, according to the value found from the global fit. 
The results are reported in Table \ref{tab:spatial24}. 
In Figure \ref{fig:Tprofile24} we report the temperature profile of A2124
in annuli around the cluster centre. 
Looking at the figure it can be seen that the temperature is not constant, but
shows a decrease with the radius increase.
\\
In order to explore if there is an asymmetry of the temperature distribution,
we divided the cluster map in four sectors 
as shown in Figure \ref{fig:sector_24}. The cluster A2122 belongs to the 
sector I, annulus 6--8 arcmin, and, unfortunately, part of its emission 
falls under the strongback.
The results of the analysis are reported in Table \ref{tab:sector24}. 
Note that in the 8$'$--12$'$
annulus, sector I, the source counts are 28.9 \% of the total counts, 
while in the remaining sectors of the same annulus the source counts are lower
than 16\%, that is the background contribution becomes dominant.
In Figure \ref{fig:sectorTprofile24} the temperature profiles in the four
sectors are presented. 
For the 0--2 arcmin region a circular region was used to fit data.
\\
Because of a shift between the X-ray centroid and the MECS centre, the
strongback support falls partly in the 6$'$--8$'$ annulus: this fact causes 
a loss of counts in sectors I and IV, which produces bad constraints in the
temperature determination, seen as exceedingly high 90\% errors (see Table
\ref{tab:sector24}).

%
%
%
\begin{table}
\caption[]{Spectral results from regions of various radii (arcmin) from the
centre of
A2124. Errors are at 90 \% and 68 \% confidence level. Abundance is fixed to the
value 0.29.}
\begin{flushleft}
\begin{tabular}{rrr}
\hline\noalign{\smallskip}
Radius ($arcmin$)& kT ($keV$) & Reduced $\chi^{2}$ (d.o.f.)\\
\noalign{\smallskip}
\hline\noalign{\smallskip}
0 - 2  & $5.44^{+0.71 \; +0.41}_{-0.61 \; -0.38}$
       & $0.81 \; (73)$ \\ 
2 - 4  & $4.24^{+0.48 \; +0.28}_{-0.41 \; -0.26}$
       & $1.07 \; (81)$ \\
4 - 6  & $3.40^{+0.58 \; +0.39}_{-0.46 \; -0.29}$
       & $0.91 \; (66)$ \\
6 - 8  & $2.82^{+1.40 \; +0.76}_{-0.75 \; -0.47}$
       & $1.10 \; (40)$ \\       
\noalign{\smallskip}
\hline
\end{tabular}
\end{flushleft}
\label{tab:spatial24}
\end{table}
%
\begin{table}
\caption[]{Spectral results from regions of various radii (arcmin) from the
centre of
A2124. Errors are at 90 \% and 68 \% confidence level. Abundance is fixed to the
value 0.29. Sectors indicated in italic are those where the source counts are
lower than 30\% of the total.}
\begin{flushleft}
\begin{tabular}{rrrr}
\hline\noalign{\smallskip}
R ($arcmin$)& Sector & kT ($keV$) & Reduced $\chi^{2}$ (d.o.f.)\\
\noalign{\smallskip}
\hline\noalign{\smallskip}
2 - 4  & I
       & $3.67^{+0.85 \; +0.51}_{-0.58 \; -0.37}$
       & $1.77 \; (27)$ \\ 
~~~~~~ & II
       & $5.10^{+1.67 \; +0.91}_{-1.06 \; -0.69}$
       & $0.52 \; (26)$ \\
~~~~~~ & III
       & $6.12^{+2.31 \; +1.26}_{-1.29 \; -0.79}$
       & $1.59 \; (28)$ \\ 
~~~~~~ & IV
       & $4.30^{+1.16 \; +0.65}_{-0.80 \; -0.52}$
       & $1.06 \; (32)$ \\  
\noalign{\smallskip}
\hline\noalign{\smallskip}                  
4 - 6  & I
       & $3.19^{+0.99 \; +0.56}_{-0.66 \; -0.43}$
       & $1.02 \; (20)$ \\ 
~~~~~~ & II
       & $3.34^{+1.21 \; +0.66}_{-0.76 \; -0.50}$
       & $1.33 \; (20)$ \\
~~~~~~ & III
       & $2.96^{+1.45 \; +0.82}_{-0.81 \; -0.56}$
       & $0.74 \; (18)$ \\ 
~~~~~~ & IV
       & $4.14^{+2.27 \; +1.17}_{-1.16 \; -0.78}$
       & $0.56 \; (17)$ \\
\noalign{\smallskip}
\hline\noalign{\smallskip}                    
6 - 8  & I
       & $2.89^{+97.11 \; +6.78}_{-99.07 \; -1.39}$
       & $0.15 \; (82)$ \\ 
~~~~~~ & II
       & $2.16^{+3.46 \; +1.42}_{-0.91 \; -0.61}$
       & $1.52 \; (12)$ \\
~~~~~~ & III
       & $2.26^{+1.56 \; +0.79}_{-0.76 \; -0.50}$
       & $0.94 \; (15)$ \\ 
~~~~~~ & IV
       & $6.41^{+ 56.64 \; +9.52}_{-3.60 \; -2.72}$
       & $0.26 \; (12)$ \\ 
\noalign{\smallskip}
\hline\noalign{\smallskip}                   
8 - 12 & {\it I}
       & ${\it 4.14^{+2.20 \; +0.92}_{-1.47 \; -1.01}}$
       & ${\it 1.29 \; (37)}$ \\
\noalign{\smallskip}
\hline
\end{tabular}
\end{flushleft}
\label{tab:sector24}
\end{table}

\section{Discussion and conclusions}

\begin{figure}
\centering
\includegraphics[angle=0,width=0.85\hsize]{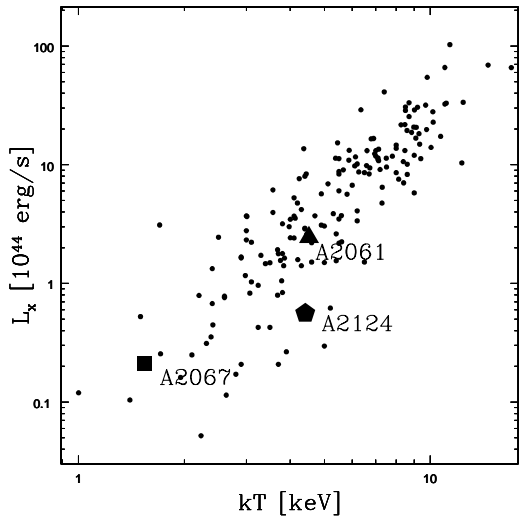}
\caption{$L_X$--$T$ relation as determined from \citet{xiang99} for a
subsample of 167 clusters. The triangle represents the location of the cluster
A2061, the square corresponds to A2067 and the pentagon is for A2124.
}
\label{fig:lumin_T}
\end{figure}

The aim of this work has been to study the gas temperature distribution of two
pairs of clusters (A2061--A2067 and A2122--A2124) located in the inner part of 
the Corona Borealis supercluster, using Beppo-SAX observations.
\\
Although the emission of A2122 falls under the window support of the detector,
in the region between the centres of A2122 and A2124
we did not find any evidence of interaction. 
Also for the pair A2061--A2067 we did not find any clear sign of
interaction: on one side A2061 is significantly elongated towards A2067, but on 
the other hand
the velocity difference $\Delta v$ between the two clusters is too large to be 
entirely due to peculiar velocity.
In order to have an estimate of the relative peculiar motion we assumed a simple
model for the collapse of two bodies approaching from infinity. We derived the
masses of A2061 and A2067 using the $M-kT$ relation \citet{yee03},
finding $v_{pec} \simeq$ 390 km s$^{-1}$, i.e. only 25\% of the $\Delta v$,
therefore the distance between A2061 and A2067 is too large to have significant
effects on the intracluster medium.
\\
In Table \ref{tab:general_information} we summarize some information 
about these clusters obtained during our study. 
\\
In Figure \ref{fig:lumin_T} we show the $L_{X}-T$ relation for a sample of 167
clusters from \citet{xiang99}, adding our data for A2061 (triangle), A2067 
(square) and A2124 (pentagon).
\\

\begin{figure*}
\centering
\includegraphics[angle=0,width=\hsize]{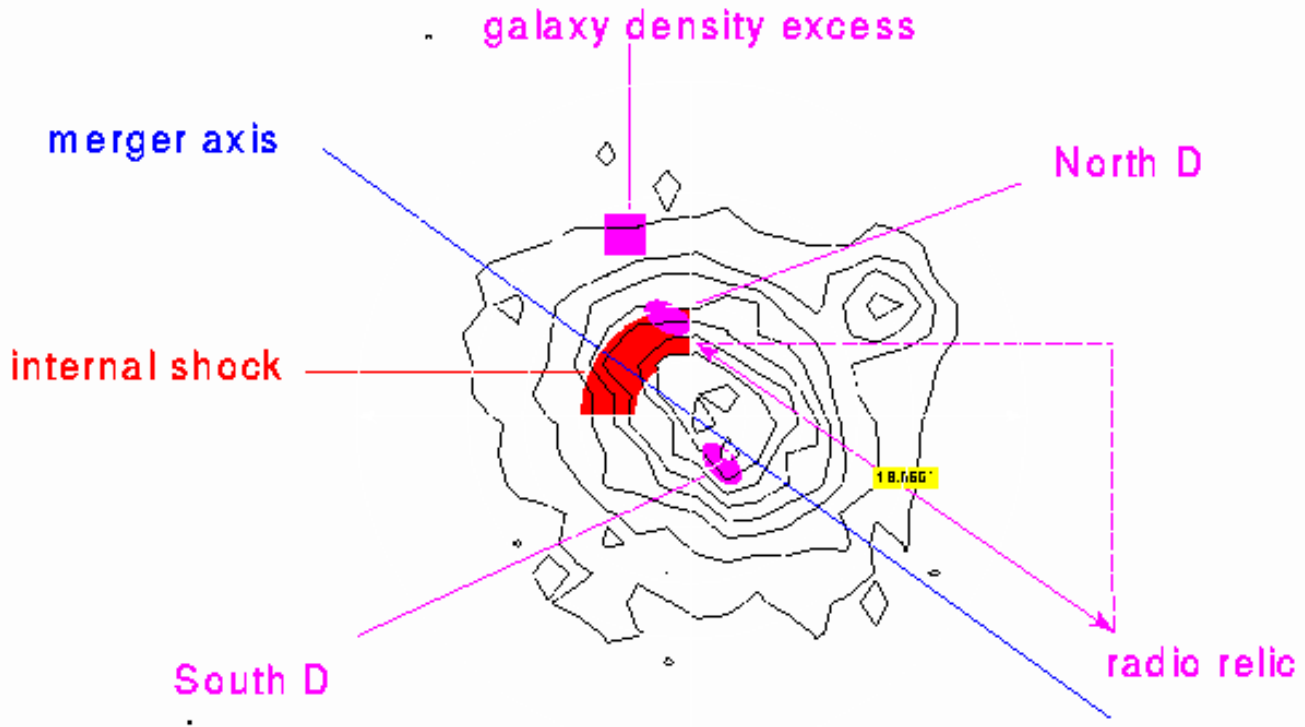}
\caption{Plan of the cluster A2061. See the text for the labelled objects}
\label{fig:cartoon_A2061}
\end{figure*}

Looking at this figure it can be seen that all the studied clusters are in
agreement with the relation. It can also be noticed that A2067 is more similar 
to a galaxy group than to a cluster, given its low temperature and its 
position in the $L_{X}-T$ plane.
\\
Analysing temperature profiles and maps of these clusters we found that their 
temperature distributions do not show strong deviations from an isothermal
behaviour. The only exception is represented by a region in A2061: we found 
that, in the bin 2--4 arcmin sector I, there is a significant increase of 
temperature ($kT$=10.67$^{+3.90}_{-2.47}$ keV) which is not 
consistent at 2.5 sigma with that
of the global fit ($kT$=4.52$^{+0.28}_{-0.24}$ keV). 
This feature likely corresponds to the presence of an internal shock.
\\
Our scenario is that a group is falling inside
A2061 and forms the shock in the major cluster. The infall is 
along the axis individuated by the elongation of the cluster 
($\sim 116^{\circ}$), as shown in Figure \ref{fig:cartoon_A2061}, that 
summarizes the geometry projected on the plane of the sky. 
Note that in the same direction, but on the opposite side, at
$\sim$ 18 arcmin ($\sim$ 1.05 \hmpc) from the X-ray centre, a relic is 
located, a radio source normally associated with an ongoing merging 
\citep{kempner01}.
However, the spatial resolution of Beppo-SAX is not able to separate the
emission of the infalling group from that of A2061. 
\\
This hypothesis is consistent with the behaviour of the two dominant galaxies:
the North D galaxy is at rest in the velocity space with respect to the
cluster, but is significantly offset with respect to the
centre of the X-ray emission (at a projected distance of $\sim$ 0.19 \hmpc).
At a projected distance of $\sim$ 0.16 \hmpc from the X-ray centre there is
another dominant galaxy, which has a velocity difference with respect to the 
cluster of
$\Delta v \sim$ 980 km s$^{-1}$. It is important to note that both the dominant 
galaxies axis inclinations ($\sim 149^{\circ}$ and $\sim 
132^{\circ}$, respectively) are similar to that of the
merger axis.
\\
By studying the bidimensional distribution of optical galaxies we found a
significant galaxy excess in the position reported as a square in Figure 
\ref{fig:cartoon_A2061}. It is likely that these galaxies
are part of the infalling group.
\\
We suppose that the group and A2061 are in the phase of interaction in which 
the cores have not 
encountered yet: in this phase the formation of a shock between the cores is
expected.
If the gas within the shocked region 
is nearly isothermal, we can calculate the ratio between the velocities after 
($u_{0}$) and before ($u_{1}$) the shock (Markevitch et al. 1999, equation (2))
as:
$$
\kappa \equiv  \frac{u_{1}}{u_{0}}   \cong  0.41 \pm 0.03 \; ,
$$
where we used the temperatures $T_{0}=4.65$ keV and $T_{1}=10.67$ keV. 
We used for  $T_{0}$ the value derived for sector II, annulus 2--4 arcmin,
which is specular to the shock position with respect to the cluster center.
\\
Given this value, the expected surface brightness ratio is:
$$
\frac{S_{X1}}{S_{X0}} \sim 6 \; .
$$
Actually we find that the ratio between the counts in the bin 2--4 arcmin 
sector IV (position of the shock) and
II (at a specular position with respect to the centre of the cluster)
is nearly 1. This inconsistency is probably due to the fact that the region
interested by the interaction is much smaller than the analysed bin and than
the MECS PSF. From simulations, produced with the same characteristics of
our observations, we estimated that this shock could be detected only if
its sizes exceed $\sim 30 \times 70$ \hkpc. 
\\
A higher spatial resolution of observation (f.i. with Chandra or XMM)
is needed to assess this scenario.
\\
The resulting difference of the gas flow after and before the shock is
(Markevitch et al. 1999, equation (1)):
$$
v_{col}  =  2 \cdot (u_{0}-u_{1}) \cong  2896^{+804}_{-588}\; 
{\rm km} \, {\rm s}^{-1}
$$
where we assume a plasma with mean molecular weight $\mu=0.60$. Note that this 
value
of $v_{col}$ is similar to that of the cluster A3667 \citep{markevitch99}.
\\
On the other hand, if we assume that the kinetic energy is completely converted 
into thermal energy, following \citet{shibata99}, we find:
$$
v_{col}=\Big [\frac{3k (T_{1} - T_{0})}{\mu m_{p}} \Big]^{1/2} \cong 
1735^{+399}_{-293} \; {\rm km} \, {\rm s}^{-1}  \; .
$$
\\
These two values found for $v_{col}$, arising from different physical
hypotheses, can be considered as upper and lower limits.
\\
This velocity range corresponds to a Mach number in the range
$$
2.4<M<3.9 \; .
$$
These values can be compared with the typical value of $M \sim$ 1.4 given by
the semi-analytical models by \citet{gabici03}: however, considering the
Mach number distribution of mergers happened in the last one billion years,
the same authors (see their figure 8) show a significant tail at $M\sim 3$.
\\
In conclusion, the global scenario can be the following.
\\
A group of galaxies, arriving from North--East, impacted on the cluster A2061. 
As a consequence of the interaction of these two structures, the galaxies of 
the group precede its intracluster medium, as predicted by numerical simulation 
(see e.g. Tormen et al. 2003).
Indeed the X-ray emission excess seen in Figure \ref{fig:ROSATimage} 
North--East of A2061 (labelled as Plume) could be the remnant of the group 
ICM, while its galaxies can be located
in correspondence of the optical overdensity shown in Figure 
\ref{fig:cartoon_A2061}.
\\
The fact that the merging axis points towards A2067 can be an indication of the
existence of a filament, on which both A2061 and A2067 lie and along which the
group merged in A2061.
%
\begin{table*}
\caption[]{ General information about A2061, A2067 and A2124. C$_{bi}$ and 
S$_{bi}$ are taken from Oegerle \& Hill (2001), while the other quantities are
derived from our analysis (see the text). Errors on temperatures 
are at the $90\%$ confidence level. }
\begin{flushleft}
\begin{tabular}{rrrrrrr}
\hline\noalign{\smallskip}
Cluster & kT$_{global}$ & L$_{bol}$ & 
F$_{[2-10]keV}$ & L$_{[2-10]keV}$ 
& C$_{bi}$    & S$_{bi}$ \\
        & keV           & erg s$^{-1}$ & erg cm$^{-2}$ s$^{-1}$ & 
erg s$^{-1}$ &  km s$^{-1}$  &   km s$^{-1}$\\
\noalign{\smallskip}
\hline\noalign{\smallskip}
Abell 2061 & 4.52$^{+0.48}_{-0.38}$ & 2.45 $\times 10^{44}$ h$^{-2}$ 
& 1.72 $\times 10^{-11}$ & 1.21 $\times 10^{44} h^{-2}$
& 23699 $\pm$ 70 &
780 $^{+57}_{-47}$\\
Abell 2067 & 1.54$^{+0.26}_{-0.22}$ & 0.21 $\times 10^{44}$ h$^{-2}$ 
& 8.62 $\times 10^{-13}$ & 0.58 $\times 10^{43} h^{-2}$ 
& 22166 $\pm$ 79 &
536 $^{+69}_{-90}$\\
Abell 2124 & 4.41$^{+0.38}_{-0.33}$ & 0.56 $\times 10^{44}$ h$^{-2}$ 
& 0.58 $\times 10^{-11}$ & 0.28 $\times 10^{44} h^{-2}$
& 19684 $\pm$ 110 &
862 $^{+91}_{-69}$\\ 
\noalign{\smallskip}
\hline
\end{tabular}
\end{flushleft}
\label{tab:general_information}
\end{table*}
%

\section*{Acknowledgments}
This research has made use of linearized event files produced
at the Beppo-SAX Science Data center.
This work has been partially supported by the Italian Space Agency grants
ASI-I-R-037-01 and ASI-I-R-063-02, and by the Italian Ministery (MIUR)
grant COFIN2001 ``Clusters and groups of galaxies: the interplay between 
dark and baryonic matter". 
This research has made use of the NASA/IPAC Extragalactic Database (NED) which 
is
operated by the Jet Propulsion Laboratory, California Institute of Technology, 
under contract with the National Aeronautics and Space Administration.
This research has made use of the SIMBAD database, operated at CDS, Strasbourg,
France.
We thank the referee (dr. H.Ebeling) for helpful comments.


 
\bsp

\label{lastpage}


\begin{thebibliography}{99}

\bibitem[\protect\citeauthoryear{Abell}{1958}]{abell58}
 Abell G.O., 1958, ApJS, 3, 211

\bibitem[\protect\citeauthoryear{Abell et al.}{1989}]{aco}
 Abell G.O., Corwin H.G.Jr., Olowin R.P., 1989, ApJS, 70, 1

\bibitem[\protect\citeauthoryear{Bahcall}{1992}]{bahcall92}
 Bahcall J.N., 1992, in Clusters and Superclusters of Galaxies, ed. A. Fabian 
 (Dordrecht: Kluwer), 275
 
\bibitem[\protect\citeauthoryear{Bardelli et al.}{1996}]{bardelli96}
 Bardelli S., Zucca E., Malizia A., Zamorani G., Scaramella R., Vettolani G.,
 1996, A\&A, 305, 435
 
\bibitem[\protect\citeauthoryear{Bardelli et al.}{1998}]{bardelli98}
 Bardelli S., Zucca E., Zamorani G., Vettolani G., Scaramella R.,
 1998, MNRAS, 296, 599

\bibitem[\protect\citeauthoryear{Bardelli et al.}{2002}]{bardelli02}
 Bardelli S., Venturi T., Zucca E., De Grandi S., Ettori E., Molendi
 S., 2002, A\&A, 396, 65
  
\bibitem[\protect\citeauthoryear{Blakeslee \& Metzger}{1999}]{blakeslee99}
 Blakeslee J.P., Metzger M.R., 1999, ApJ, 513, 592

\bibitem[\protect\citeauthoryear{Boella et al.}{1997a}]{boella97a}
 Boella G., Butler R.C., Perola G.C., Piro L., Scarsi L., Bleeker
 J.A.M., 1997a, A\&AS, 122, 299
 
\bibitem[\protect\citeauthoryear{Boella et al.}{1997b}]{boella97b}
 Boella G.,  et al., 1997b, A\&AS, 122, 327
 
\bibitem[\protect\citeauthoryear{Cappi \& Maurogordato}{1992}]{cappi92}
 Cappi A., Maurogordato S., 1992, A\&A, 259, 423  
 
\bibitem[\protect\citeauthoryear{Cavaliere \& FuscoFemiano}{1976}]{cavaliere76}
 Cavaliere A., Fusco-Femiano R., 1976, A\&A, 49, 137
 
\bibitem[\protect\citeauthoryear{D'Acri et al.}{1998}]{dacri98} 
 D'Acri F., De Grandi S., Molendi S., 1998, Nucl.Phys., 69/1-3, 581
 
\bibitem[\protect\citeauthoryear{De Grandi \& Molendi}{2001}]{degrandi01} 
 De Grandi S., Molendi S., 2001, ApJ, 551, 153
 
\bibitem[\protect\citeauthoryear{De Grandi \& Molendi}{2002}]{degrandi02} 
 De Grandi S., Molendi S., 2002, ApJ, 567, 163 
  
\bibitem[\protect\citeauthoryear{Dickey \& Lockman}{1990}]{dickey90} 
 Dickey J.M., Lockman F.J., 1990, ARA\&A, 28, 215 
 
\bibitem[\protect\citeauthoryear{Ebeling et al.}{1996}]{ebeling96}
 Ebeling H., Voges W., B\"{o}hringer H., Edge A.C., Huchra J.P., Briel
 U.G., 1996, MNRAS, 281, 799
 
\bibitem[\protect\citeauthoryear{Edge et al.}{1990}]{edge90}
 Edge A.C., Stewart G.C., Fabian A.C., Arnaud K.A., 1990, MNRAS, 245, 599 
 
\bibitem[\protect\citeauthoryear{Ettori et al.}{1997}]{ettori97}
 Ettori S., Fabian A.C., White D.A., 1997, MNRAS, 289, 787
 
\bibitem[\protect\citeauthoryear{Ettori \& Fabian}{1999}]{ettori99}
 Ettori S.,  Fabian A.C., 1999, MNRAS, 305, 843 
 
\bibitem[\protect\citeauthoryear{Ettori et al.}{2000}]{ettori00}
 Ettori S., Bardelli S., De Grandi S., Molendi S., Zamorani G., Zucca
 E., 2000, MNRAS, 318, 239
  
\bibitem[\protect\citeauthoryear{Fiore et al.}{1999}]{fiore99} 
 Fiore F., Guainazzi M., Grandi P., 
 1999, Cookbook for Beppo-SAX NFI Spectral Analysis
 
\bibitem[\protect\citeauthoryear{Gabici \& Blasi}{2003}]{gabici03}
 Gabici S., Blasi P., 2003, ApJ, 583, 695 

\bibitem[\protect\citeauthoryear{Kaastra}{1992}]{kaastra92} 
 Kaastra J.S.,  1992, An X-Ray Spectral Code for Optically Thin Plasma, 
 Internal SRON-Leiden report, updated version 2.0
 
\bibitem[\protect\citeauthoryear{Kempner \& Sarazin}{2001}]{kempner01}
 Kempner J.C., Sarazin C.L., 2001, ApJ, 548, 639 
 
\bibitem[\protect\citeauthoryear{Kopylova \& Kopylov}{1998}]{kopylov98} 
 Kopylova F.G., Kopylov A.I., 1998, AstL, 24, 4 
 
\bibitem[\protect\citeauthoryear{Lubin \& Bahcall}{1993}]{lubin93} 
 Lubin L.M., Bahcall, N.A., 1993, ApJ, 415, L17 
 
\bibitem[\protect\citeauthoryear{Markevitch et al.}{1999}]{markevitch99}
 Markevitch M., Sarazin C.L., Vikhlinin A., 1999, ApJ, 521, 526 
 
\bibitem[\protect\citeauthoryear{McKee et al.}{1980}]{mckee80}
 McKee J.D., Mushotzky R.F., Boldt E.A., Holt S.S., Marshall F.E.,
 Pravdo S.H., Serlemitsos P.J., 1980, ApJ, 242, 843
  
\bibitem[\protect\citeauthoryear{Mewe}{1995}]{mewe95} 
 Mewe R., Kaastra J., Liedhal K., 1995, Legacy 6, 16
 
\bibitem[\protect\citeauthoryear{Oegerle \& Hill}{1998}]{oegerle98}
 Oegerle W.R., Hill J.M., 1998, AJ, 116, 1529 
 
\bibitem[\protect\citeauthoryear{Oegerle \& Hill}{2001}]{oegerle01}
 Oegerle W.R., Hill J.M., 2001, ApJ, 122, 2858 
 
\bibitem[\protect\citeauthoryear{Oppenheimer et al.}{1997}]{oppenheimer97}
 Oppenheimer B.R., Helfand D.J., Gaidos E.J., 1997, AJ, 113, 2134
 
\bibitem[\protect\citeauthoryear{Parmar et al.}{1997}]{parmar97} 
 Parmar A.N., et al., 1997, A\&AS, 122, 309
 
\bibitem[\protect\citeauthoryear{Perri \& Capalbi}{2002}]{perri02}
 Perri M., Capalbi M., 2002, A\&A, 396, 759 
 
\bibitem[\protect\citeauthoryear{Postman et al.}{1988}]{postman88}
 Postman M., Geller J.M., Huchra J.P., 1988, AJ, 95, 267 
 
\bibitem[\protect\citeauthoryear{Sarazin}{1988}]{sarazin88}
 Sarazin C.L., 1988, X-ray emission from the cluster of galaxies, Cambridge
 University Press 
 
\bibitem[\protect\citeauthoryear{Sarazin}{2000}]{sarazin00}
 Sarazin C.L., 2000, in Costructing the universe with Clusters of Galaxies, 
 Durret F. \& Gerbal D. eds., electronic proceedings
 http://www.iap.fr/Conferences/Colloque/coll2000/contributions
 
\bibitem[\protect\citeauthoryear{Shibata et al.}{1999}]{shibata99}
 Shibata R., Honda H., Ishida M., Ohashi T., Yamashita K., 1999, ApJ,
 524,603

\bibitem[\protect\citeauthoryear{Small et al.}{1997}]{small97}
 Small T.A., Sargent W.L.W., Hamilton D., 1997, ApJS, 111, 1
 
\bibitem[\protect\citeauthoryear{Small et al.}{1998}]{small98}
 Small T.A., Ma C., Sargent W.L.W., Hamilton D., 1998, ApJ, 492, 45
 
\bibitem[\protect\citeauthoryear{Struble \& Rood}{1999}]{struble99}
 Struble M.F., Rood H.J., 1999, ApJS, 125, 35
 
\bibitem[\protect\citeauthoryear{Tormen}{1997}]{tormen97} 
 Tormen G., 1997, MNRAS, 290, 411  
 
\bibitem[\protect\citeauthoryear{Tormen}{2004}]{tormen04} 
 Tormen G., Moscardini L., Yoshida N., 2004, MNRAS, 350, 1397

\bibitem[\protect\citeauthoryear{Venturi et al.}{2000}]{venturi00}
 Venturi T., Bardelli S., Morganti R., Hunstead R.W., 2000, MNRAS,
 314, 594
 
\bibitem[\protect\citeauthoryear{Wue et al.}{1999}]{xiang99}
 Wue X., Xue Y., Fang L., 1999, ApJ, 407, 470 
 
\bibitem[\protect\citeauthoryear{Yee \& Ellingson}{2003}]{yee03}
 Yee H.K.C., Ellingson E., 2003, ApJ, 585, 215 
 
\bibitem[\protect\citeauthoryear{Zucca et al.}{1993}]{zucca93}
 Zucca E., Zamorani G., Scaramella R., Vettolani G., 1993, ApJ, 407,
 470
    
\end{thebibliography}
\end{document}